\documentclass[galaxies,article,accept,moreauthors,pdftex,10pt,a4paper]{mdpi}
\firstpage{1}
\articlenumber{x}
\doinum{10.3390/------}
\pubvolume{xx}
\pubyear{2016}
\copyrightyear{2016}
\externaleditor{Academic Editors: Jose L. G\'{o}mez, Alan P. Marscher and Svetlana G. Jorstad}
\history{Received: 5 September 2016; Accepted: 25 October 2016; Published: date}

\usepackage{graphicx}
\usepackage{caption}
\usepackage{subcaption}
\usepackage{soul}

\Title{Physical Conditions and Variability Processes in AGN Jets through Multi-Frequency Linear and Circular Radio Polarization Monitoring}

\Author{Ioannis Myserlis *, Emmanouil Angelakis, Alex Kraus, Lars Fuhrmann, Vassilis Karamanavis and J. Anton Zensus}
\AuthorNames{Ioannis Myserlis, Emmanouil Angelakis, Alex Kraus, Lars Fuhrmann, Vassilis Karamanavis and J. Anton Zensus}

\address[1]{Max-Planck-Institut f\"ur Radioastronomie, Auf dem H\"ugel 69, 53121 Bonn, Germany; imyserlis@mpifr-bonn.mpg.de; Tel.: +49-228-525-389}

\abstract{Radio polarimetry is an invaluable tool to investigate the physical conditions and variability processes in  active galactic nuclei (AGN) jets. However, detecting their linear and circular polarization properties is a challenging endeavor due to their low levels and possible depolarization effects. We have developed an end-to-end data analysis methodology to recover the polarization properties of unresolved sources with high accuracy. It has been applied to recover the linear and circular polarization of 87 AGNs measured by the F-GAMMA program from July 2010 to January 2015 with a mean cadence of 1.3 months. Their linear polarization was recovered at four frequencies between 2.64 and 10.45~GHz and the circular polarization at 4.85 and 8.35~GHz. The physical conditions required to reproduce the observed polarization properties and the processes which induce their variability were investigated with a full-Stokes radiative transfer code which emulates the synchrotron emission of modeled jets. The model was used to investigate the conditions needed to reproduce the observed polarization behavior for the blazar 3C\,454.3, assuming that the observed variability is attributed to evolving internal shocks propagating downstream.}


\begin{document}





\section{Introduction/Overview}
The jets of active galactic nuclei (AGN) contain low density, relativistic plasmas and magnetic fields, e.g., \citep{Begelman1984}. Those two constituents are responsible for the synchrotron component of their emission which spans from the radio to the optical, UV or even X-rays and is intrinsically both linearly and circularly polarized. The polarization parameters (linear and circular polarization degrees, circular polarization handedness and polarization angle) carry information about the physical conditions in the regions where the radiation is generated and propagated through, such as the magnetic field strength and topology, the particle density and the plasma composition, both in the jet as well as any magnetized plasma regions along the line of sight.

The radio emission of AGN jets often shows pronounced variability. The variability is usually attributed to propagating shocked regions in the jet, where the magnetic field is compressed, resulting in an increase of the polarization degree, e.g., \citep{Marscher1985,Hughes1985,Hughes1989}. The optical depth evolution following the initial compression changes the observed polarization characteristics \citep{Marscher1985,Myserlis2014}, offering an insight into the evolution of the physical conditions at the emission site during such periods of increased variability.

However, the detection of their polarized components, especially the circular one, is challenging due to the low polarization levels and possible depolarizing effects like the complex source structure or the tangledness of the jet's magnetic field. Single-dish observations often yield linear polarization degree measurements of about 3\%--5\% and circular polarization degree measurements of about 0.5\%, with the later rarely reaching levels as high as 1\%--2\%, e.g., \citep{Klein2003,Myserlis2015}.

We have developed an end-to-end polarimetric data analysis methodology for the Effelsberg~100-m telescope, which eliminates a number of systematics bringing the uncertainty to levels as low as 0.1\%, allowing the detection of the inherently low polarization degrees. \mbox{The methodology} was applied to recover multi-frequency polarization properties for 87 AGN jets, monitored with the F-GAMMA program \citep{Fuhrmann2016,Fuhrmann2007,Angelakis2010} between July 2010 and January 2015.

Finally, we developed a full-Stokes radiative transfer code, based on Hughes et al.  \citep{Hughes1989}, which emulates the synchrotron emission of a modeled jet. The variability is induced by the propagation of relativistic shock fronts along the jet axis. The model was used to reproduce the total flux and polarization variability observed in the prototype blazar 3C\,454.3 between MJD
 $\sim$55400 and $\sim$55800. A number of physical properties were constrained, such as the jet plasma density, the coherence length of the magnetic field as well as the compression and Doppler factors of the propagating shocked regions. Our results are in agreement with independent previous estimates in the literature.

\section{High-Precision Linear and Circular Polarimetry with the 100-m Telescope}
We developed an end-to-end polarimetric data analysis methodology which can be used to recover the linear and circular polarization properties of unresolved sources in the radio window. The methodology was developed using data obtained with the 4.85-GHz and 8.35-GHz receivers of the 100-m Effelsberg telescope, equipped with circularly polarized feeds. It eliminates a number of systematics bringing the uncertainty to levels as low as 0.1\% for linear polarization degree, 0.5$^{\circ}$ for polarization angle and 0.2\% for circular polarization degree measurements. The most important features of our methodology are:

\subsection{Instrumental Linear Polarization Correction}
The observing system introduces spurious signals in the receiver channels responsible for the Stokes $Q$ and $U$ measurements. Those signals are the manifestation of the instrumental linear polarization and they can either (a) induce false polarization in unpolarized sources, (b) obscure weak polarization signals present or (c) modify the polarization signal of polarized sources.

We use the Stokes $Q$ and $U$ datasets obtained from linearly unpolarized sources to create a model of the instrumental linear polarization signals for every observing session. This model is then used to recreate the expected shape and amplitude of the instrumental polarization signals which are subtracted from each measurement.

\subsection{Optimization of Beam Pattern Fitting Model}
The observables, i.e., the amplitude, the full width at half maximum and the peak offset of the telescope's response for each measurement are extracted by a fitting operation. We investigated different beam pattern models in order to optimize this procedure. Our results show that the Airy disk pattern, i.e. the diffraction pattern of a circular aperture describes the whole area of the dataset with high accuracy. This approach is essential for the inherently low Stokes $V$ measurements ($\sim$0.5\%) using circularly polarized feeds.

\subsection{Instrumental Circular Polarization Correction}
The instrumental circular polarization is manifested by a systematic offset of the Stokes $V$ measurements. Under the assumption that the instrumental circular polarization is caused by an imbalance between the gains of the receiver channels sensitive to the left and right circularly polarized components of the incident radiation, we use the Stokes $V$ measurements of sources which are expected to have stable circular polarization to restore the gain balance. These sources are either (a) emitters of circularly unpolarized radiation, such as the planetary nebula NGC\,7027 which is a free-free emitter or (b) sources which are unlikely to vary, such as the steep spectrum sources 3C\,286 and 3C\,48. For the later we don't  need to assume a certain degree or handedness of circular~polarization.

\begin{figure}[H]
    \centering
    \begin{subfigure}[t]{0.5\textwidth}
        \centering
        \includegraphics[height=10cm]{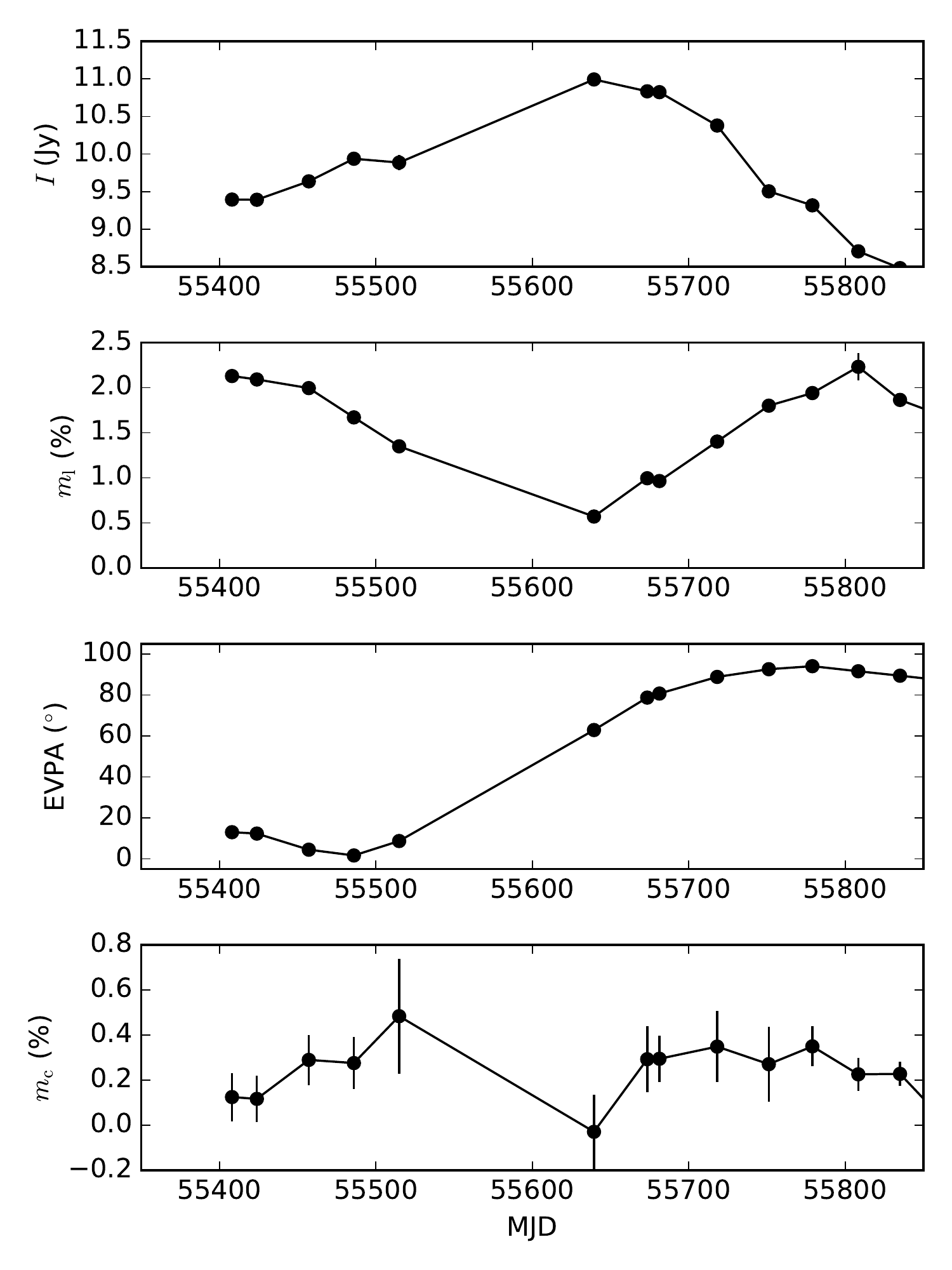}
        \caption{}
    \end{subfigure}%
    \begin{subfigure}[t]{0.5\textwidth}
        \centering
        \includegraphics[height=10cm,trim={0 0 4.8cm 0},clip]{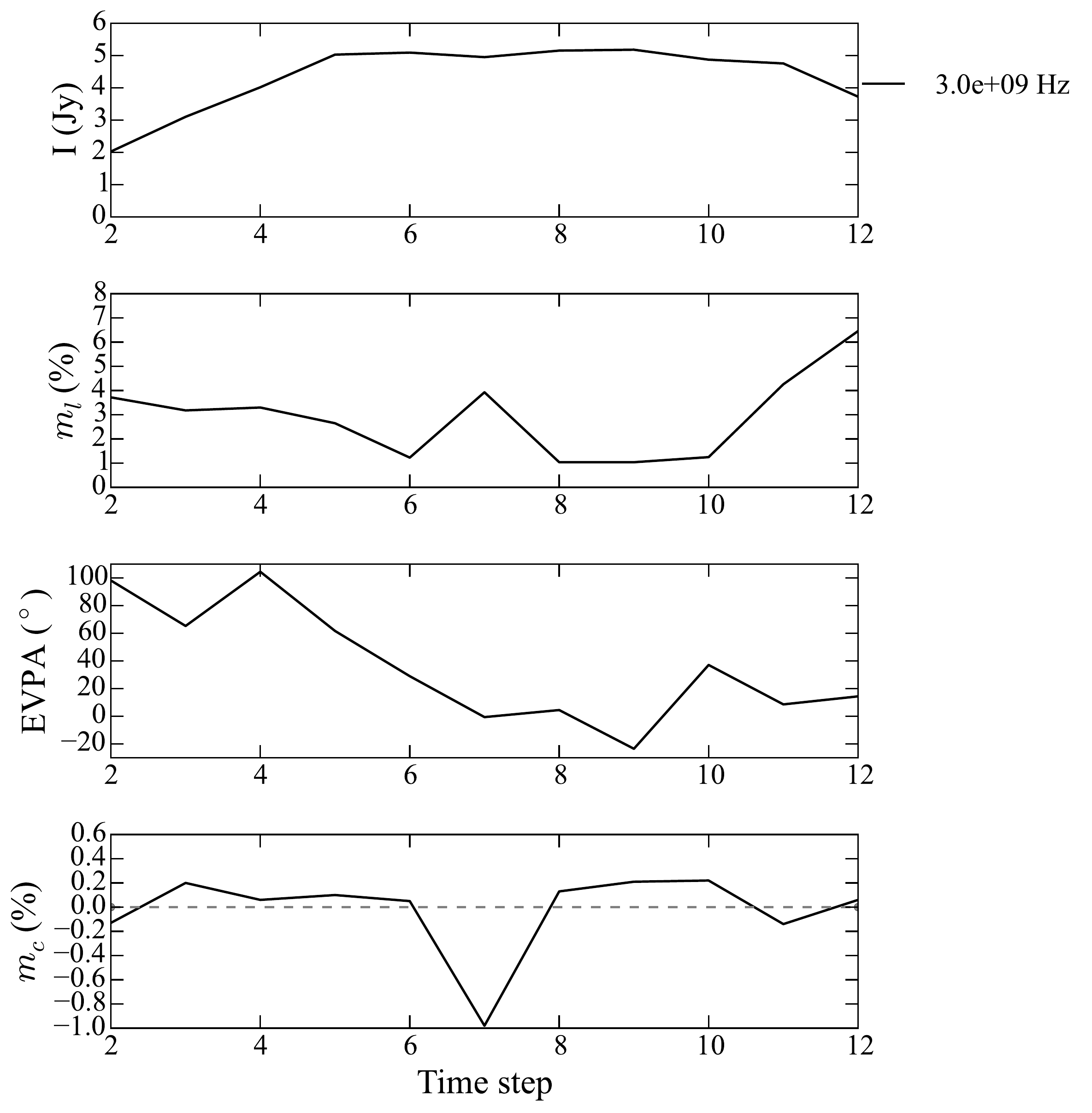}
        \caption{}
    \end{subfigure}
    \vspace{-12pt}
    \caption{\textbf{(a)} The 8.35-GHz total flux density and polarization lightcurves of the blazar 3C\,454.3 for the MJD
     range of $\sim$55400 to $\sim$55800. From top to bottom: (1) Stokes $I$; (2) the degree of linear polarization $m_{\mathrm{l}}$; (3) the polarization angle (EVPA) and (4) the degree of circular polarization $m_{\mathrm{c}}$. \textbf{(b)} A reproduction of the 8.35-GHz total flux density and polarization variability (left). The synthetic full-Stokes lightcurves were generated by our~model.}
    \label{fig:454_example}
\end{figure}

The methodology which is briefly described above was used to recover the linear and circular polarization parameters of 87 AGN jets, monitored with the F-GAMMA program with a mean cadence of 1.3 months \citep{Fuhrmann2016,Fuhrmann2007,Angelakis2010}. Until now, we have optimized the methodology to recover the linear polarization parameters at 2.64, 4.85, 8.35 and 10.45~GHz and the circular polarization at 4.85\linebreak  and 8.35~GHz. This resulted in the construction of full-Stokes light curves from July 2010 to \mbox{January 2015} for the polarized sources in our sample. An example is shown in Figure~\ref{fig:454_example}a. In the future, we plan to extend the linear and circular polarization data sets both in frequency \mbox{(2.64--14.6 GHz)} and in time (2007--2015).

\section{Probing the Physical Conditions in the Jet of 3C\,454.3}
The radio variability of the jets we monitored often follows repeating patterns in the flux density--frequency domain which agree with the predictions of the ``shock-in-jet'' model \citep{Marscher1985}. For a number of sources, including the blazar 3C\,454.3, our observations revealed also the coordinated changes of the polarization characteristics which mark the transitions between the optically thick and thin regimes of synchrotron emission \citep{Pacholczyk1977,Myserlis2014}. Assuming that these transitions are due to the optical depth evolution of shocks propagating downstream, we developed a full-Stokes radiative transfer model, based on the description in Hughes et al. \citep{Hughes1989}, in order to investigate the physical conditions of the jet in both the shocked and unshocked parts of the flow.

\subsection{Model Overview}
\begin{figure}[H]
\centering
\includegraphics[trim={10 100 10 100}, clip, width=0.9\hsize]{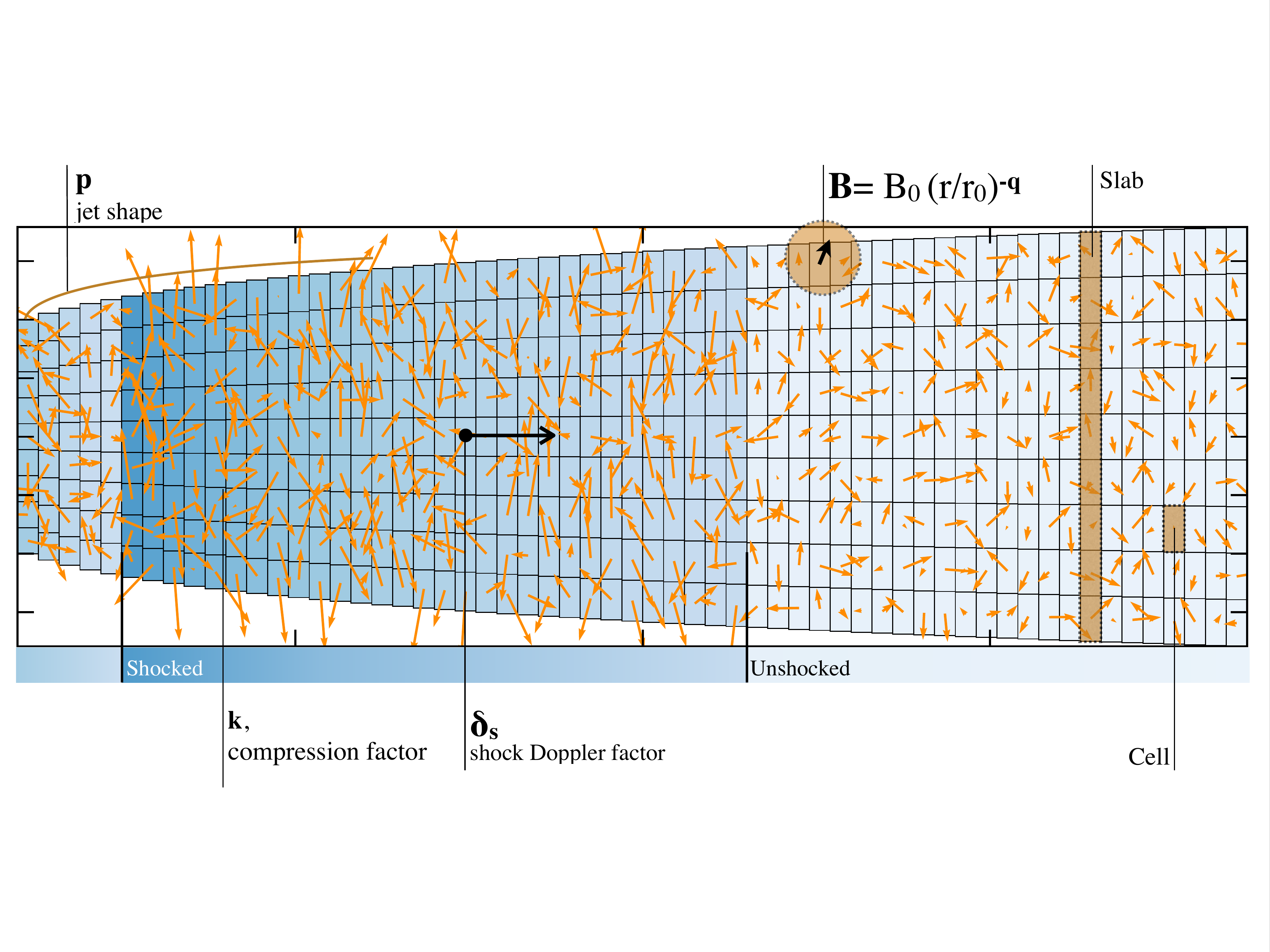}
\caption{A modeled jet profile and some key parameters of the radiative transfer code. The density gradient along the jet axis is demonstrated by the color shading of the cells. The orange arrows show the projection of magnetic field on the plane of the sky.}
\label{fig:model}
\end{figure}

We model an AGN jet as an ensemble of cells, each with homogeneous relativistic plasma and uniform magnetic field (Figure~\ref{fig:model}). Our model emulates the emission in terms of all four Stokes parameters ($I$, $Q$, $U$ and $V$), for both the unshocked and shocked regions of the plasma in the jet using a set of known jump conditions, valid for relativistic shock fronts. The variability is induced by the relativistic propagation of a disturbance along the jet axis, in the form of a shock (dark blue area in Figure~\ref{fig:model}). This shock has a certain compression factor $k$ which changes the local physical conditions, such as the density, the low energy cutoff and the magnetic field strength. Those changes are then imprinted on the emitted spectrum which follows the shock evolution, i.e., the propagation of a Synchrotron Self-Absorbed (SSA) component from high towards lower frequencies.

The spectra of the modeled jet at any given time step are obtained by solving the full-Stokes radiative transfer problem for radiation emitted and propagated across the jet  using the equations given in Jones  and Odell \citep{Jones1977a} and Hughes et al. \citep{Hughes1989}. According to these solutions, the mechanisms which generate or modify the polarized emission are the synchrotron radiation, Faraday rotation and Faraday conversion. Synthetic total flux and polarization lightcurves are then generated by sampling the full-Stokes evolving spectra at pre-selected radio frequencies. The direct comparison of the synthetic and the observed lightcurves allows us to constrain a number of physical conditions for both the shocked and unshocked flow regions.

\subsection{3C\,454.3: A Case Study}

\subsubsection{Linear and Circular Polarization Variability}
The multi-frequency radio variability of the blazar 3C\,454.3 exhibits repeating patterns in the flux density--frequency domain which can be attributed to the propagation of SSA components through the observing band-pass, as predicted by the ``shock-in-jet'' model \citep{Marscher1985}. According to synchrotron theory, if the emitting region is threaded with a uniform magnetic field, we expect to observe the following set of characteristic changes as the SSA components move from the optically thick to the thin regime:
\begin{enumerate}\renewcommand\labelenumi{(\theenumi)}
\item A minimization of the linear polarization degree, concurrent with a polarization angle rotation of exactly 90$^{\circ}$,
\item a minimization of the circular polarization degree, followed by a change in the circular polarization handedness and
\item a maximization of the total flux when the peak of the SSA spectrum matches the observing~frequency.
\end{enumerate}
Two of the above changes were found in the polarimetric dataset of 3C\,454.3 between MJD $\sim$55400 and $\sim$55800 as shown in Figure~\ref{fig:454_example}a. A polarization angle (EVPA) rotation of exactly 90$^{\circ}$ was observed around MJD $\sim$55650 (Figure~\ref{fig:454_example}a, 3rd panel), concurrent with a minimization of the linear polarization degree (2nd panel). Those changes are also followed by a maximization of the total flux density (1st panel) as predicted by the above description. The circular polarization degree and handedness  seem to remain stable around that time (4th panel).

\subsubsection{Variability Modeling and Constrained Parameters}
The full-Stokes radiative transfer code was used to reproduce the total flux and polarization variability observed in the blazar 3C\,454.3 between MJD $\sim$55400 and $\sim$55800. A comparison between the 8.35-GHz observed and synthetic light curves from our model is shown in Figure~\ref{fig:454_example}. In both cases we see a maximization of the total flux density, a minimization of the linear polarization degree, a rotation of exactly 90$^{\circ}$ in the polarization angle while the circular polarization degree shows a relatively stable behavior. The physical conditions we managed to constrain by exploring the parameter space of our model were:
\begin{itemize}[leftmargin=*,labelsep=5.5mm]
\item the coherence length of the jet's magnetic field, which is in principle equal to the cell size of our model, to $\sim$9 pc,
\item the jet plasma density to $\sim$10--100 cm$^{-3}$,
\item the compression factor of the propagating shock, $k \approx 0.8$, and
\item the Doppler factor of the propagating shock, $D \approx 30$, in agreement with independent previous estimates in the literature \citep{Hovatta2009,Sasada2014,Zhou2015}.
\end{itemize}

\section{Discussion/Conclusions}
We developed an end-to-end data analysis pipeline to recover the linear and circular polarization properties of point-like sources in the radio window. Our analysis eliminates a number of systematics bringing the uncertainty to levels as low as 0.1\% for linear polarization degree, 0.5$^{\circ}$ for polarization angle and 0.2\% for circular polarization degree measurements. We used our pipeline to recover the polarization properties of 87 AGNs measured by the F-GAMMA program \citep{Fuhrmann2016,Fuhrmann2007,Angelakis2010} from July 2010 to\linebreak \mbox{January 2015} with a mean cadence of 1.3 months. Their linear polarization was recovered at 2.64, 4.85, 8.35 and 10.45~GHz and the circular polarization at 4.85 and 8.35~GHz.

For a number of sources the observed total flux and polarization variability can be attributed to the ``shock- in-jet'' model \citep{Marscher1985,Myserlis2014}. We developed a full-Stokes radiative transfer model, based on the work of Hughes et al. \citep{Hughes1989}, which emulates the synchrotron emission of a modeled jet.\linebreak The variability is induced by the propagation of relativistic shock fronts along the jet axis. The model can be used to reproduce the total flux and polarization variability of the observed sources based on the predictions of the ``shock-in-jet'' model. Here we present our first results on modeling the variability of the prototype blazar 3C\,454.3 between MJD $\sim$55400 and $\sim$55800. By searching the parameter space of our model, we managed to constrain a number of physical conditions such as the jet plasma density, the coherence length of the magnetic field as well as the compression and Doppler factors of the propagating shocked regions. Our results are in agreement with independent previous estimates in the literature.

\newpage


\acknowledgments{This research is based on observations with the 100-m telescope of the MPIfR (Max-Planck-Institut f\"ur Radioastronomie) at Effelsberg. I.M. and V.K. were funded by the International Max Planck Research School (IMPRS) for Astronomy and Astrophysics at the Universities of Bonn and Cologne. I.M. thanks  W. Max-Moerbeck, the internal MPIfR referee, for useful comments on this paper.}

\authorcontributions{E.A. and I.M. conceived and designed the project; I.M., E.A. and V.K. performed the observations and analyzed the data; A.K, L.F. and J.A.Z. contributed in the analysis and interpretation of the data; I.M. wrote the paper.}

\conflictofinterests{The authors declare no conflict of interest. }
\vspace{-12pt}

\bibliographystyle{mdpi}

\begin{thebibliography}{999}

\bibitem[Begelman \em{et~al.}(1984)Begelman, Blandford, and Rees]{Begelman1984}
Begelman, M.C.; Blandford, R.D.; Rees, M.J.
\newblock {Theory of extragalactic radio sources}.
\newblock {\em Rev. Modern Phys.} {\bf 1984}, {\em 56},~255--351.

\bibitem[Marscher and Gear(1985)]{Marscher1985}
Marscher, A.P.; Gear, W.K.
\newblock {Models for high-frequency radio outbursts in extragalactic sources,
  with application to the early 1983 millimeter-to-infrared flare of 3C 273}.
\newblock {\em Astrophys. J.} {\bf 1985}, {\em 298},~114--127.

\bibitem[Hughes \em{et~al.}(1985)Hughes, Aller, and Aller]{Hughes1985}
Hughes, P.A.; Aller, H.D.; Aller, M.F.
\newblock {Polarized Radio Outbursts in Bl-Lacertae - Part Two - the Flux and
  Polarization of a Piston-Driven Shock}.
\newblock {\em Astrophys. J.} {\bf 1985}, {\em 298},~301.

\bibitem[Hughes \em{et~al.}(1989)Hughes, Aller, and Aller]{Hughes1989}
Hughes, P.A.; Aller, H.D.; Aller, M.F.
\newblock {Synchrotron emission from shocked relativistic jets. I - The theory
  of radio-wavelength variability and its relation to superluminal motion}.
\newblock {\em Astrophys. J.} {\bf 1989}, {\em 341},~54--79.

\bibitem[Myserlis \em{et~al.}(2014)Myserlis, Angelakis, Fuhrmann, Pavlidou,
  Nestoras, Karamanavis, Kraus, and Zensus]{Myserlis2014}
Myserlis, I.; Angelakis, E.; Fuhrmann, L.; Pavlidou, V.; Nestoras, I.;
  Karamanavis, V.; Kraus, A.; Zensus, J.A.
\newblock {Multi-frequency linear and circular radio polarization monitoring of
  jet emission elements in Fermi blazars}.
\newblock  {\bf 2014}, arXiv:1401.2072.

\bibitem[{Klein} \em{et~al.}(2003){Klein}, {Mack}, {Gregorini}, and
  {Vigotti}]{Klein2003}
{Klein}, U.; {Mack}, K.H.; {Gregorini}, L.; {Vigotti}, M.
\newblock {Multi-frequency study of the B3-VLA sample. III. Polarisation
  properties}.
\newblock {\em Astron. Astrophys.} {\bf 2003}, {\em 406},~579--592.

\bibitem[{Myserlis}(2015)]{Myserlis2015}
{Myserlis}, I.
\newblock {A framework for the study of physical conditions in astrophysical
  plasmas through radio and optical polarization---Application to
  extragalactic jets}.
\newblock PhD Thesis, Max-Planck-Institut f{\"u}r Radioastronomie, Bonn, Germany,  2015.
\newblock URI:	\url{http://kups.ub.uni-koeln.de/id/eprint/6967}

\bibitem[{Fuhrmann} \em{et~al.}(2016){Fuhrmann}, {Angelakis}, {Zensus},
  {Nestoras}, {Marchili}, {Pavlidou}, {Karamanavis}, {Ungerechts}, {Krichbaum},
  {Larsson}, {Lee}, {Max-Moerbeck}, {Myserlis}, {Pearson}, {Readhead},
  {Richards}, {Sievers}, and {Sohn}]{Fuhrmann2016}
{Fuhrmann}, L.; {Angelakis}, E.; {Zensus}, J.A.; {Nestoras}, I.; {Marchili},
  N.; {Pavlidou}, V.; {Karamanavis}, V.; {Ungerechts}, H.; {Krichbaum}, T.P.;
  {Larsson}, S.; et al.
\newblock {The F-GAMMA program: Multi-frequency study of Active Galactic Nuclei
  in the Fermi era. Program description and the first 2.5 years of monitoring}.
\newblock {\bf 2016}, arXiv:astro-ph.HE/1608.02580.

\bibitem[Fuhrmann \em{et~al.}(2007)Fuhrmann, Zensus, Krichbaum, Angelakis, and
  Readhead]{Fuhrmann2007}
Fuhrmann, L.; Zensus, J.A.; Krichbaum, T.P.; Angelakis, E.; Readhead, A.C.S.
\newblock {Simultaneous Radio to (Sub-) mm-Monitoring of Variability and
  Spectral Shape Evolution of potential GLAST Blazars}.
\newblock  \textit{AIP Conf. Proc.} \textbf{2007}, \textit{921}, 249--251.

\bibitem[Angelakis \em{et~al.}(2010)Angelakis, Fuhrmann, Nestoras, Zensus,
  Marchili, Pavlidou, and Krichbaum]{Angelakis2010}
Angelakis, E.; Fuhrmann, L.; Nestoras, I.; Zensus, J.A.; Marchili, N.;
  Pavlidou, V.; Krichbaum, T.P.\linebreak
\scalebox{.95}[1.0]{\newblock {The F-GAMMA program: multi-wavelength AGN studies in the Fermi-GST era}.
\newblock  {\bf 2010}, arXiv:1006.5610.}

\bibitem[Pacholczyk(1977)]{Pacholczyk1977}
Pacholczyk, A.G.
\newblock {\em Radio galaxies: Radiation Transfer, Dynamics, Stability and
  Evolution of a Synchrotron Plasmon};
\newblock Pergamon Press: Oxford, UK, 1977; p.105-109.

\bibitem[Jones and Odell(1977)]{Jones1977a}
Jones, T.W.; Odell, S.L.
\newblock {Transfer of polarized radiation in self-absorbed synchrotron sources.
  I. Results for a homogeneous source}.
\newblock {\em Astrophys. J.} {\bf 1977}, {\em 214},~522--539.

\bibitem[Hovatta \em{et~al.}(2009)Hovatta, Valtaoja, Tornikoski, and
  L{\"{a}}hteenm{\"{a}}ki]{Hovatta2009}
Hovatta, T.; Valtaoja, E.; Tornikoski, M.; L{\"{a}}hteenm{\"{a}}ki, A.
\newblock {Doppler factors, Lorentz factors and viewing angles for quasars, BL
  Lacertae objects and radio galaxies}.
\newblock {\em Astron. Astrophys.} {\bf 2009}, {\em 494},~527--537.

\bibitem[Sasada \em{et~al.}(2014)Sasada, Uemura, Fukazawa, Yasuda, Itoh,
  Sakimoto, Ikejiri, Yoshida, Kawabata, Akitaya, Ohsugi, Yamanaka, Komatsu,
  Miyamoto, Nagae, Nakaya, Tanaka, Sato, and Kino]{Sasada2014}
Sasada, M.; Uemura, M.; Fukazawa, Y.; Yasuda, H.; Itoh, R.; Sakimoto, K.;
  Ikejiri, Y.; Yoshida, M.; Kawabata, K.S.; Akitaya, H.; et al.
\newblock {Extremely high polarization in the 2010 outburst of blazar 3C
  454.3}.
\newblock {\em Astrophys. J.} {\bf 2014}, {\em 784},~141.

\bibitem[Zhou \em{et~al.}(2015)Zhou, Yan, and Dai]{Zhou2015}
Zhou, Y.; Yan, D.H.; Dai, B.Z.
\newblock {The optical variability properties of flat spectrum radio quasar 3C
  454.3}.\linebreak
\newblock {\em New Astron.} {\bf 2015}, {\em 36},~19--25.

\end{thebibliography}

\renewcommand\bibname{References}


\end{document}